\documentclass[preprint]{aastex62}
\begin{document}

\author{E. Jourdain}
\affil{ CNRS; IRAP; 9 Av. colonel Roche, BP 44346, F-31028 Toulouse cedex 4, France\\
 Universit\'e de Toulouse; UPS-OMP; IRAP;  Toulouse, France\\}
\author{J.-P. Roques}
\affil{ CNRS; IRAP; 9 Av. colonel Roche, BP 44346, F-31028 Toulouse cedex 4, France\\
 Universit\'e de Toulouse; UPS-OMP; IRAP;  Toulouse, France\\}
\title{2003-2018 Monitoring of the Crab Nebula Polarization in hard X-rays with \textit{INTEGRAL} SPI.}

\begin{abstract}
We have analyzed 16 years of observations dedicated to the Crab(pulsar  + nebula) with the \textit{INTEGRAL} SPI instrument to investigate its polarization properties. We find that the source presents a substantially polarized emission (PF = 24\%) in the hard X-ray domain, with the electric vector  aligned with the pulsar spin axis, in agreement with other results at various wavelengths. The stability of the polarization characteristics with energy and over the 16 years covered by the data is remarkable, completing the standard candle status of the source in the spectral domain. The polarization measurements imply that  the synchrotron emission is the dominant mechanism of photon production from radio to hard X-rays. The high level of polarized emission points out the steadiness of the source, in particular of the magnetic field configuration and  geometry.
\end{abstract}

\section{Introduction}
SPI is a hard X-ray/soft $\gamma$-ray spectrometer providing an excellent energy resolution in the 20 keV- 8 MeV energy range with some imaging capabilities. In addition, the design of the detector plane, with 19 independent crystals, makes possible the measurement of the polarization parameters of the incident radiation, for energies above $\sim$ 100 keV. Indeed, \citet{dean08} have analyzed the SPI data of the Crab pulsar (off-pulse emission) and reported the first detection of a polarized emission in the hard X-ray domain. Then, \citet{forot08} confirm the polarization of the Crab emission with the imager IBIS, also aboard the \textit{INTEGRAL}  mission. Since, several instruments followed the way, enlarging the investigated energy domain. In this paper, we analyzed the SPI data accumulated on  the Crab nebula since the launch of the \textit{INTEGRAL}  mission. The  large amount of data allows us to study the polarization characteristics of the source emission both over the time and as a function of the energy.

\section{Observations and Data Analysis}
In this section, we will focus on the information specifically relevant to the polarimetry study. The reader  can refer to  \citet{Vedrenne03} for an overview of the SPI instrument and to \citet{Roques03} for the in flight performance, while a description of the standard data analysis can be found in \citet{crab09} and \citet{roques19}.
  
\subsection{The Data set}
Since its launch, the \textit{INTEGRAL}  observatory performs regular observations of the Crab nebula, to allow calibration
monitoring/updates of the onboard instruments. These observations are performed twice a year,
in February-March, then in September. The first campaign in March 2003, just after the end of the mission performance validation phase, gathered 446 ks of data (useful duration; Revolutions 43-44-45). After that, the bi-annual campaigns consisted of relatively short exposures ($\sim$ 50 ks), often dedicated to peculiar configuration tests (off-axis pointings, mask corner study...). Since 2008, each calibration campaign lasts 2 revolutions, i.e. $\sim$ 400 ks, twice a year, with, in general, a standard 5X5 pattern pointing strategy. In addition, shorter observations (45 ks) are planned every 4 revolutions all along the Crab visibility periods, but these will not be considered in the following. 
Before starting our analyses,  the exposures which present unstable background or other issues (pointing anomaly, outburst of the neighboring source A0535+262, etc.) have been removed. The final data set   encompasses $\sim$ 6.4  Ms, from March 2003 to September 2018.
Further, to seek for any source evolution, the data set has been split into four periods, as a compromise between a timescale as short as possible, and an adequate signal to noise ratio. 
Due to the lack of suitable observations between March 2003 and October 2005, we consider separately the March 2003 revolutions.  The remaining dataset has been broken up into three parts of similar useful  durations (see Table \ref{LogP}).  These four sub-data sets will be referred to with their respective labels, P1 to P4, in the subsequent analyses.

\subsection{SPI as a Polarimeter}
The SPI polarimetric capacities rely on the Compton interactions of high energy photons in the detection plane. This latter consists of 19 Germanium crystals, and photons above $\sim$ 100 keV may  diffuse in a first crystal and escape toward another one, where a second interaction occurs and so on, until a photo-electric absorption or final escaping. These events are called 'multiple event' (hereafter ME). The characteristics of the energy deposits and detectors involved contain crucial information on the polarization porperties of the incident photons. In Germanium detectors, the  fraction of ME (i.e. Compton interactions with the diffused photon escaping 
toward the next detector) becomes non negligible above $\sim$ 90 keV. However, ME remain    minoritarian and  represent only $\sim$ 20\% of the total incident flux integrated above 100 keV. In fact, most of the Compton diffused photons are photo-absorbed in the same detector. This low efficiency implies long integration durations to obtain a good signal-to-noise ratio.

 In practice, we consider only photons which hit two adjacent detectors (double events or ME2). This turns out to handle 42 pseudo-detectors (42 possible pairs of adjacent detectors), instead of 19 detectors as done in the standard analysis. 
Concerning the spectra and light curve production, the standard analysis tools, used routinely for   reconstructing the incident flux from 'single detector' events (SE and PE\footnote{SE and PE flags correspond to events which loss energy in only one detector (Single detector Events). If such an event triggers a second specific electronic chain (PSD module, dedicated to a pulse shape analysis), it is flagged PE, if not, it is flagged SE. See \citet{roques19} for details}.)
%single detector energy deposit photons
%(which lost energy in one detector; note that this case include Compton interaction with photo-electric absorption inside the same detector or escape of the diffused photon), 
can be applied to ME2 events. The appropriate response matrices have been produced, together with the standard matrices \citep{sturner03}, and the flux extraction procedure is the same. To validate the ME2 fluxes which are considered in the polarimetry study described below, we  build the corresponding spectra by deconvolving the ME2 counts with relevant matrices and compare them to the single detector event spectra.

The  procedure specifically developed for the polarization  studies has been detailed in \citet{chauvin13}. The main features are:
\begin{itemize}
\item Selection of ME2 events in the 42 pseudo-detectors: Each pseudo-detector is associated  with the total energy deposit (sum of the two measured energies). 

\item Simulations of the instrument responses to a polarized emission, for 17 polarization angles (PA, from 0$^\circ$ to 170$^\circ$ by step of 10$^\circ$), and 100 polarization fractions (PF, 0\% to 100\% by step of 1\%), considering the Crab localization in the FoV, for each  pointing.

\item Comparison of simulations and observational data, for a given set of pointings: Source and background normalizations are estimated  by the resolution of an equation system for each (PA, PF) pair.
\begin{equation}
D_{sd}=x \times G4_{sd}(PF,PA)+y \times B_{sd}
\label{Source_equ}
\end{equation}
where $D_{sd}$ is the observed count distribution for a science window (or exposure) s, in the pseudo-detector d; x is the source normalisation; G4, the simulated count distribution, for the same s and d, as a function of the
source polarization fraction, PF, and angle, PA; y is the background 
normalization and B, the background spatial distribution, taken from an empty field  observation. The  simulated counts are renormalized to the corresponding detector livetimes. The x and y values are determined through a linear least-squares resolution and the resulting $\chi2$ value is stored. 
At the end of the PA and PF loops, the lower $\chi2$ between model and data identifies the best parameters.
\end{itemize}

\section{results}

\subsection{spectral analysis}
 
A spectral analysis from both single and multiple events  has been  performed, in order to compare the respective averaged spectra. 
For each of the four periods mentioned above, both spectra have been fit simultaneously. 
The Crab nebula emission has been described by the Band model (GRBM in Xspec language), proposed
   by \citet{band93} to model the GRB spectra. This analytical model  reproduces the smooth curvature observed between 20 and $\sim$ 1 MeV better than a broken power law \citep{roques19}. The same synchrotron origin of both GRB and pulsar emissions further supports this choice.
For each period, the shape parameters have been coupled between both spectra, while the individual normalizations are kept free.
\begin{figure}
\includegraphics[scale=0.8]{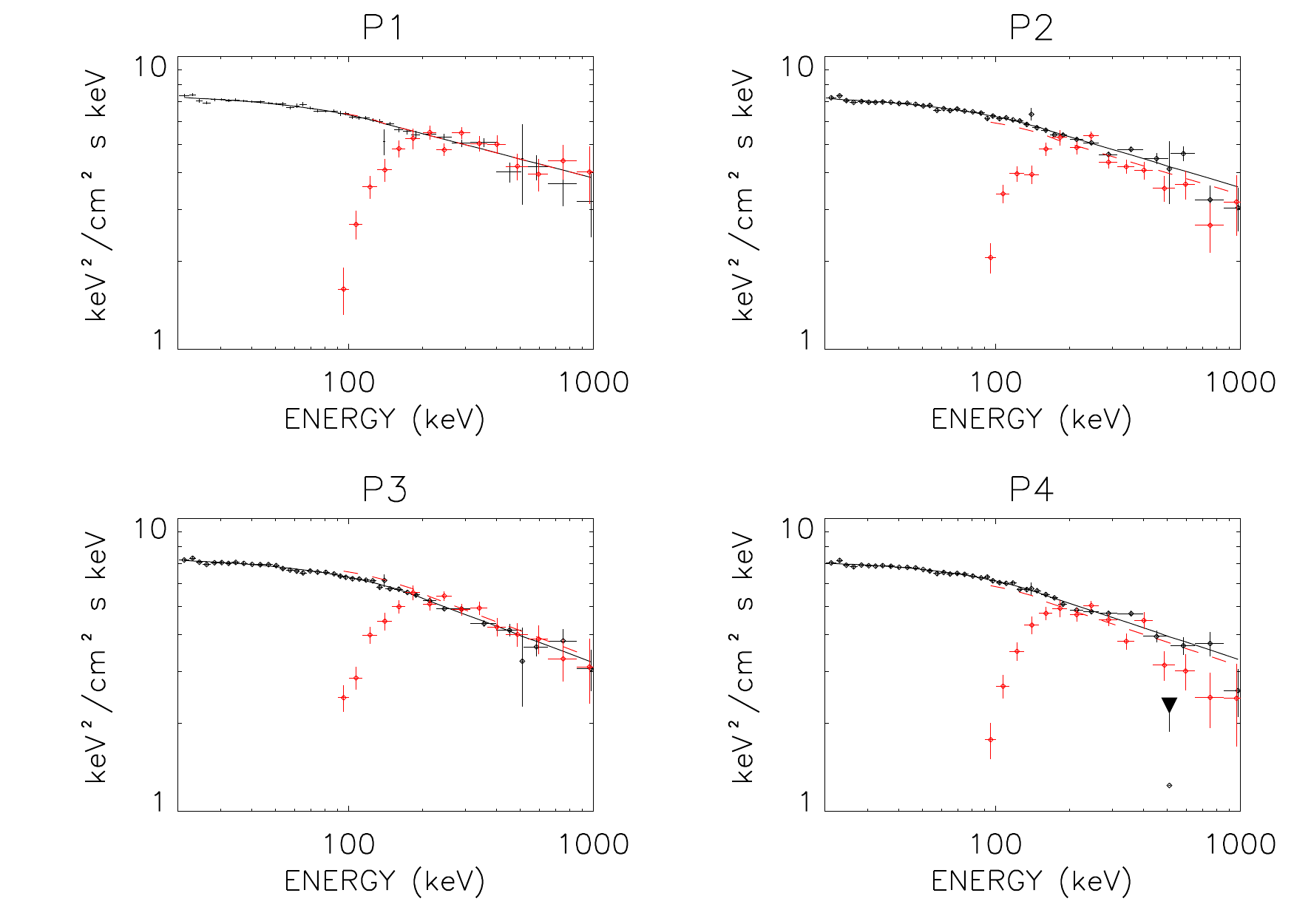} 
\caption{Spectra extracted from SE+PE (black points) and ME2 (red points) for the four periods described in Table \ref{LogP}.} \label{SP}
 \end{figure}
The  results of the spectral analyses are presented in  Table \ref{Fitgrbm} and Figure \ref{SP}. 
Furthermore,  the normalization factors (for E $\simeq$ 170 keV) agree  within 5\%  in any period. The data below $\simeq$ 170 keV suffer from the uncertainties in the ME2 efficiency embedded in the response matrices. These inaccuracies do not affect the polarization results since they have no specific anisotropy on the detector plane.
Finally, the perfect agreement between both spectra for each period demonstrates the reliability of the ME2 flux extraction.

\subsection{Polarization study}

\begin{figure}
\includegraphics[scale=0.25]{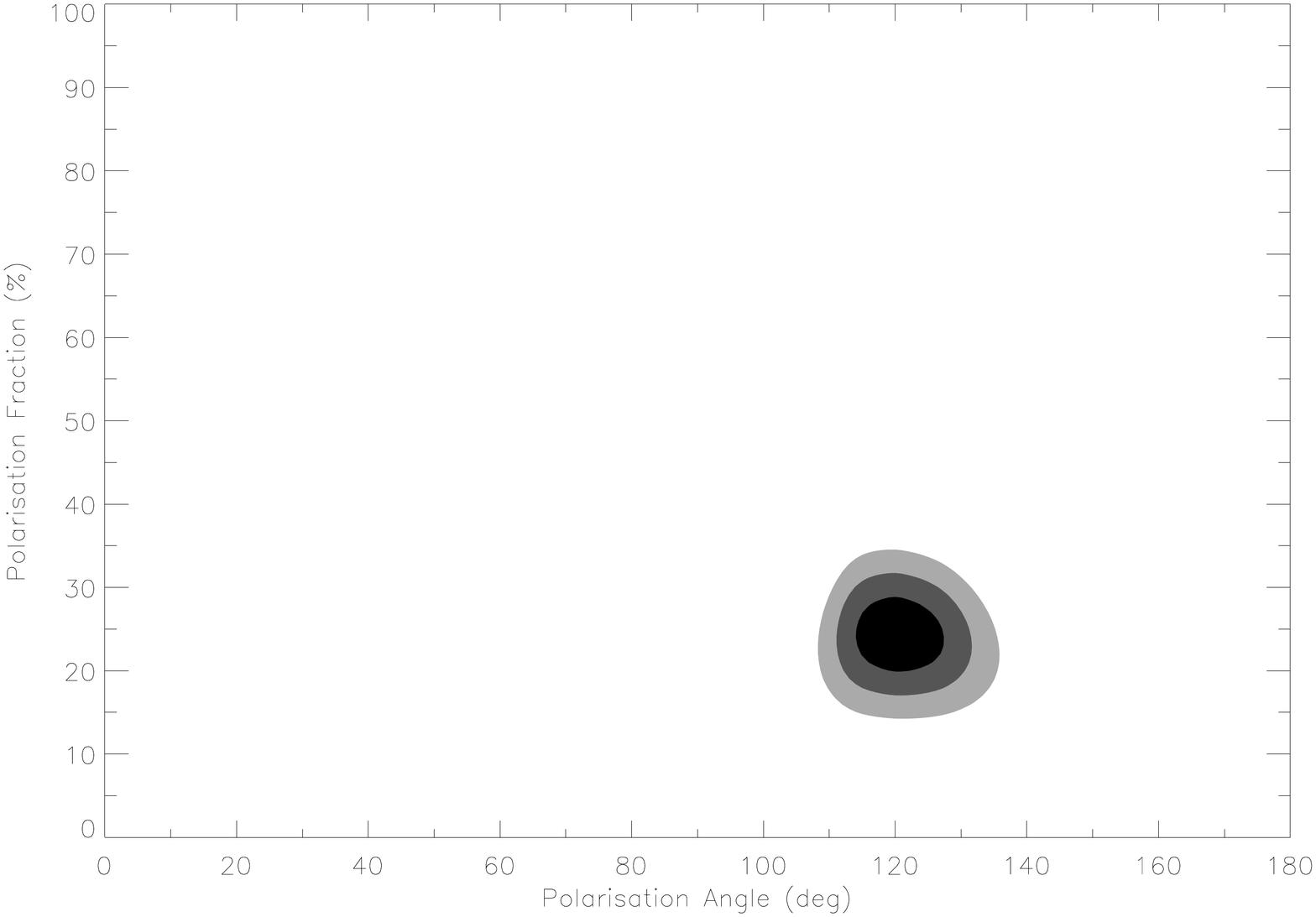}
\includegraphics[scale=0.25]{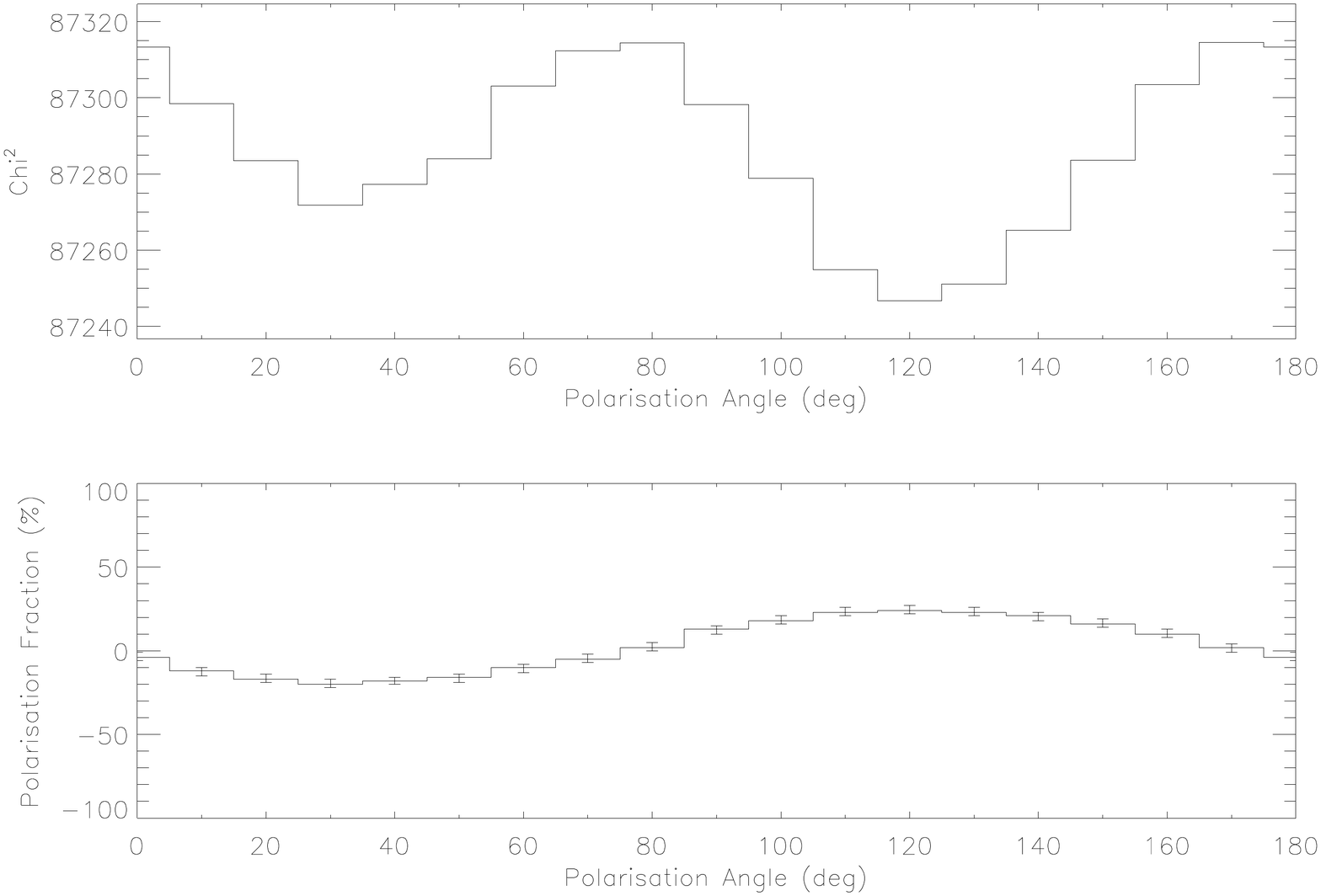}
\caption{Best-fit polarization parameters of the Crab nebula, between 130 and 436 keV, for the 6.4 Ms data set, from 2003 to 2018. Left: significance contours in the PA-PF plane, for $\chi2_{min}$+ 2.7, 6.18 and 11.8, i.e. 1, 2 and 3 $\sigma$ confidence levels. Right: Minimum $\chi2$  (top) and associated FP (bottom)  in function of PA}\label{tot}.
\end{figure}

Once the ME2 events were validated, the procedure described in the previous section has been applied to the total data set. We have selected photons with  energies  between 130 and 436 keV, in order to optimize the signal-to-noise ratio. Note that the events between 196 and 201 keV have been removed, due to the strong background line present at 198 keV.
Assuming   constant values all along the 16 years,  we obtain a Polarization Angle (PA) of 120$^\circ \pm 6^\circ $ with a Polarization Fraction (PF) of  24 $\pm 4$\%. This result is visualized in Figure \ref{tot}, in the  PA-PF  plane, with the 2-D surface contours  calculated from the  $\chi2$ map, at 
$\chi2_{min}$+ 2.7, 6.18 and 11.8, corresponding to 1, 2 and 3 $\sigma$ confidence  levels  for two free parameters. 
 
\begin{figure} 
\includegraphics[scale=0.6]{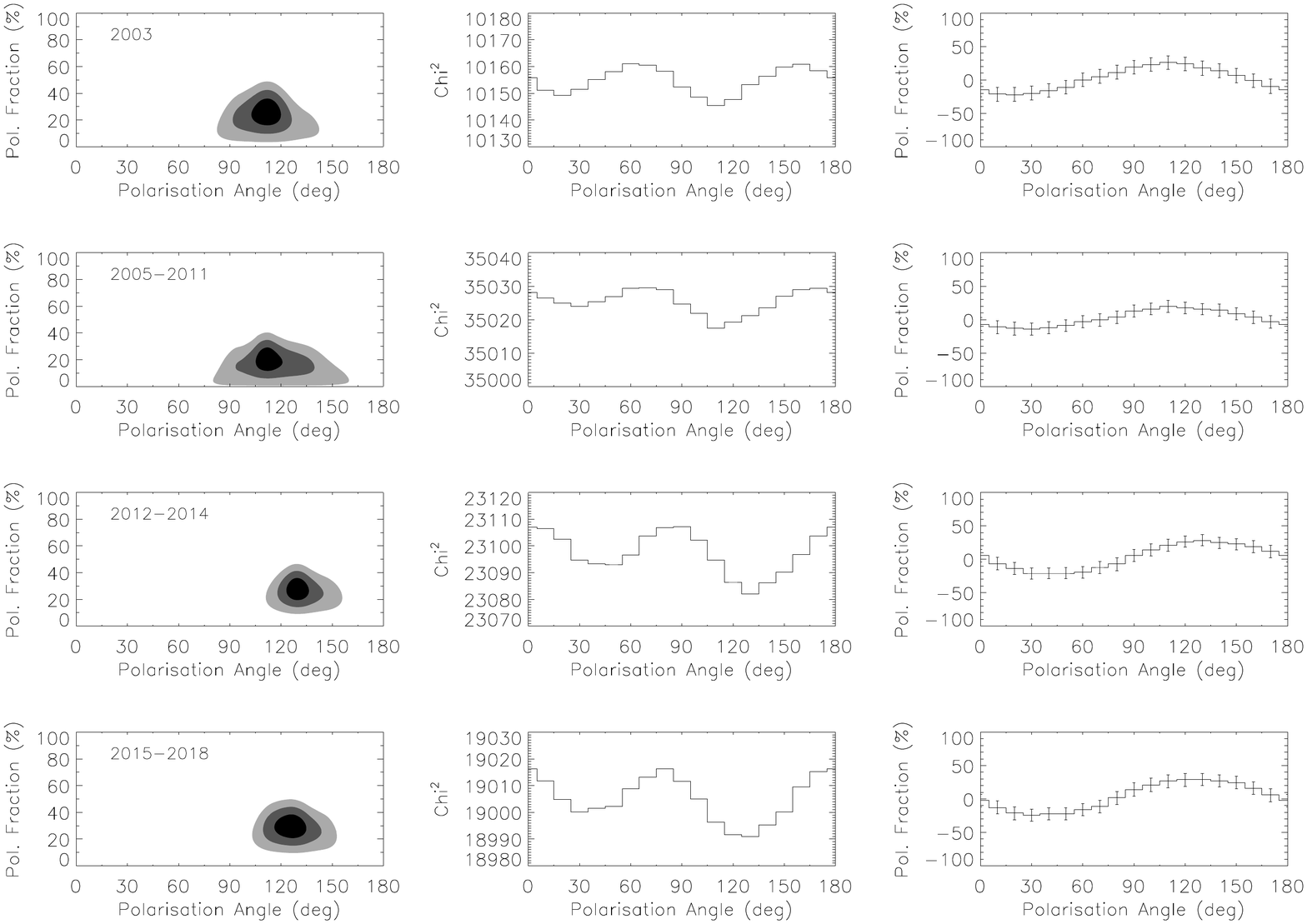} 
\caption{The same as Figure \ref{tot}, for the four periods defined in Table
\ref{LogP}.}\label{polarvstimea}
\end{figure}

\begin{figure} 
\includegraphics[scale=0.3]{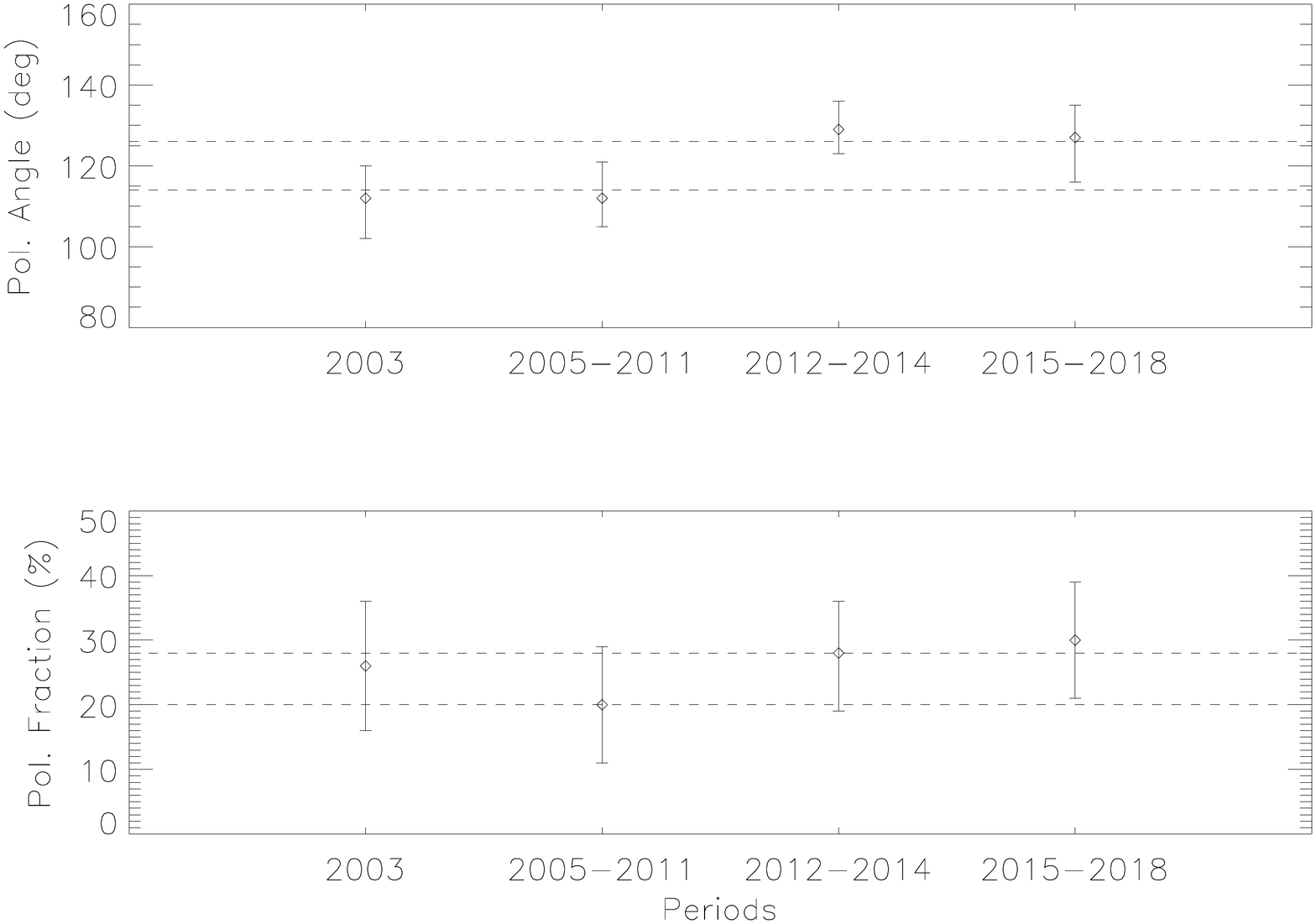} 
\includegraphics[scale=0.3]{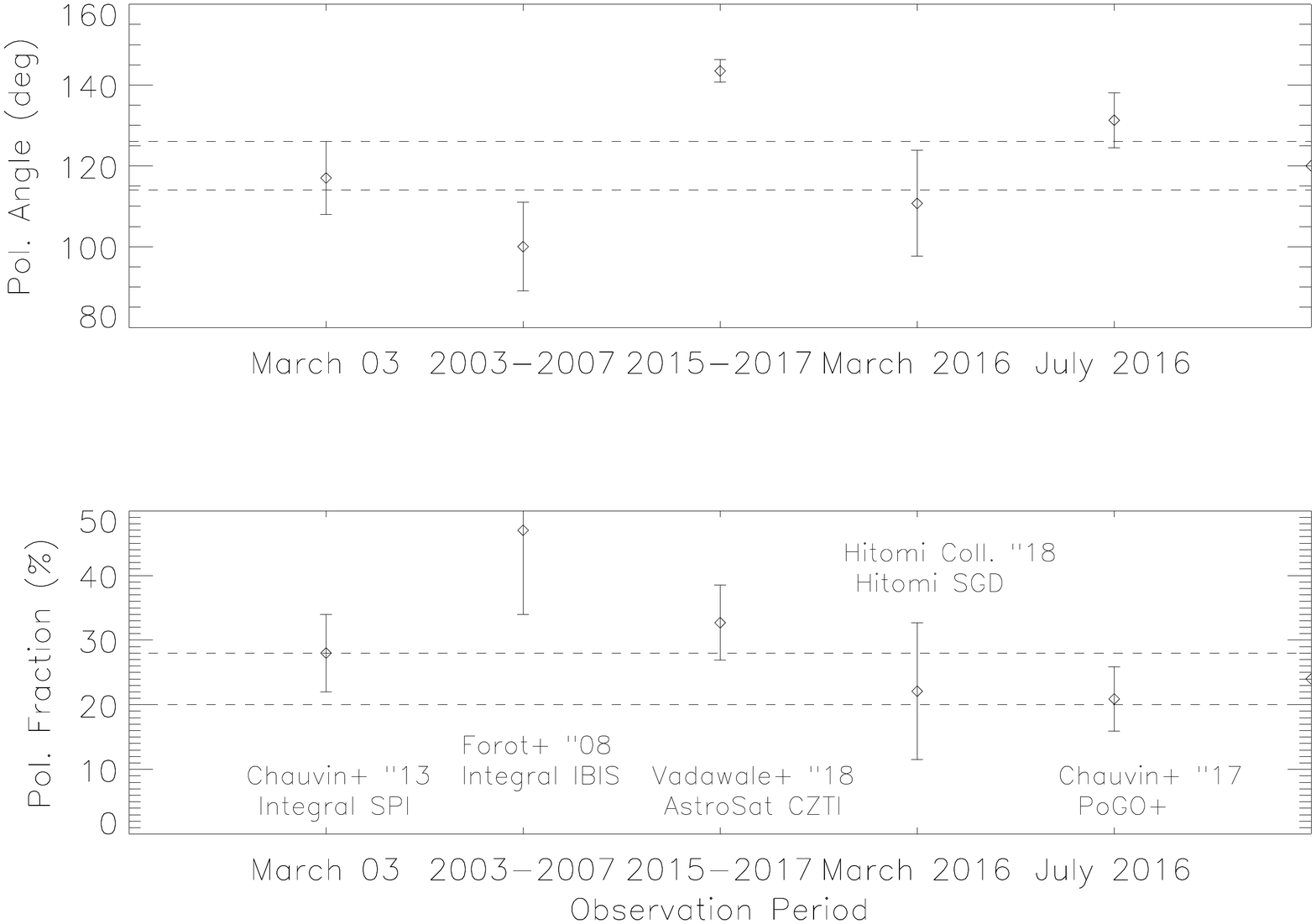} 
\caption{Left: Evolution of the SPI polarization parameters with time, in the 130-436 keV energy range. The errors quoted are 1 $\sigma$ for 2 parameters of interest. Right: Results from various instruments (references in the text).
In both panels, dashed lines represent the 1-$\sigma$ uncertainty interval for our  SPI total data set best-fit values.}\label{polarvstimeb}
\end{figure}

To verify the stability of the source over time, the same analysis has been performed separately for  periods P1 to P4. The individual results are displayed  in Figure \ref{polarvstimea}. In Figure \ref{polarvstimeb},  the evolution of the best fit polarization parameters are plotted, and compared to the mean values, obtained from the total data set (dashed lines = 1 $\sigma$ uncertainty interval). No significant evolution is visible, all values being compatible with the respective mean values.
  

\begin{figure}
\includegraphics[scale=0.6]{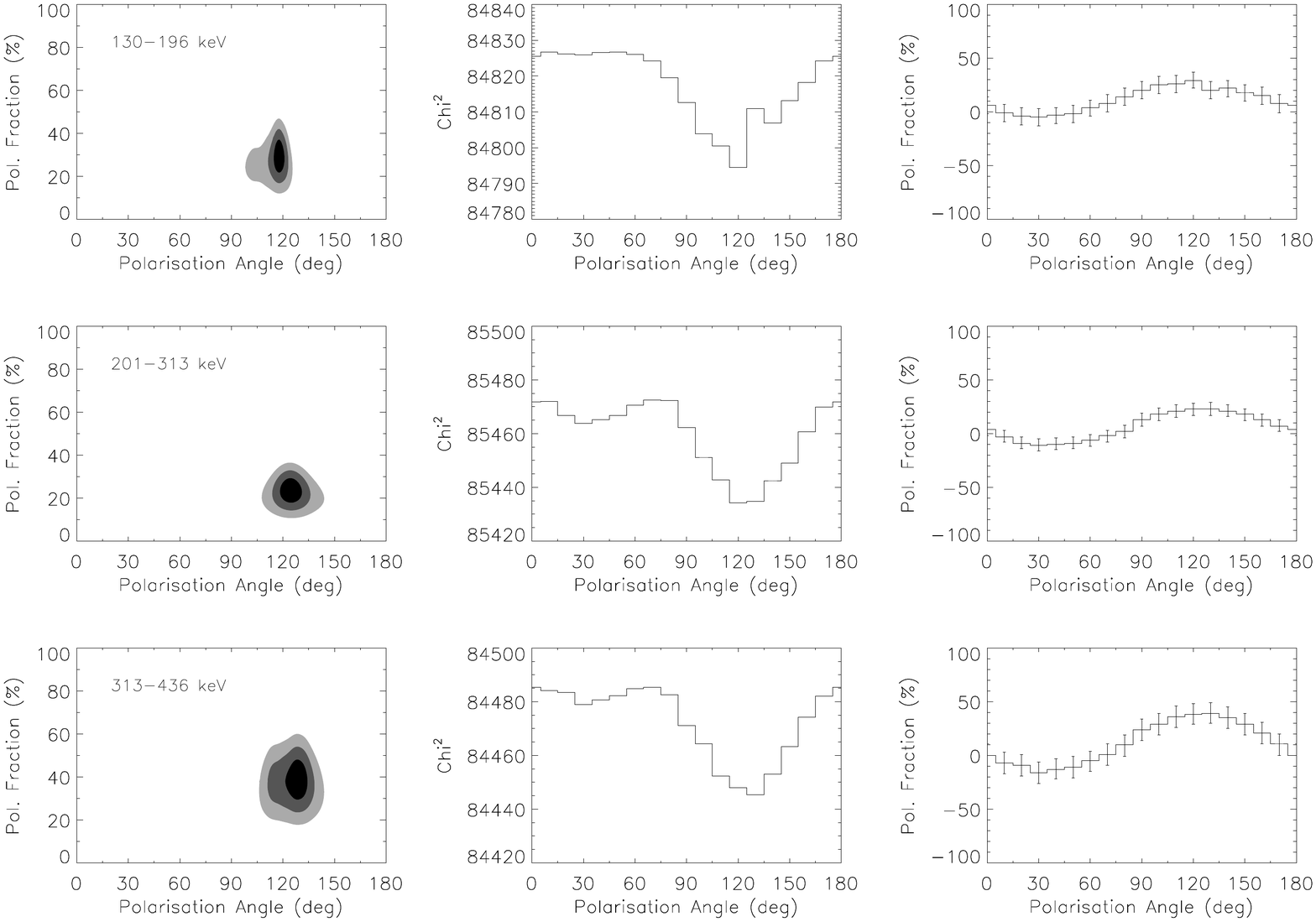}
\caption{The same as Figure \ref{tot}, for three narrower energy bands.}\label{polarvsEa}
\end{figure}

\begin{figure}
\includegraphics[scale=0.3]{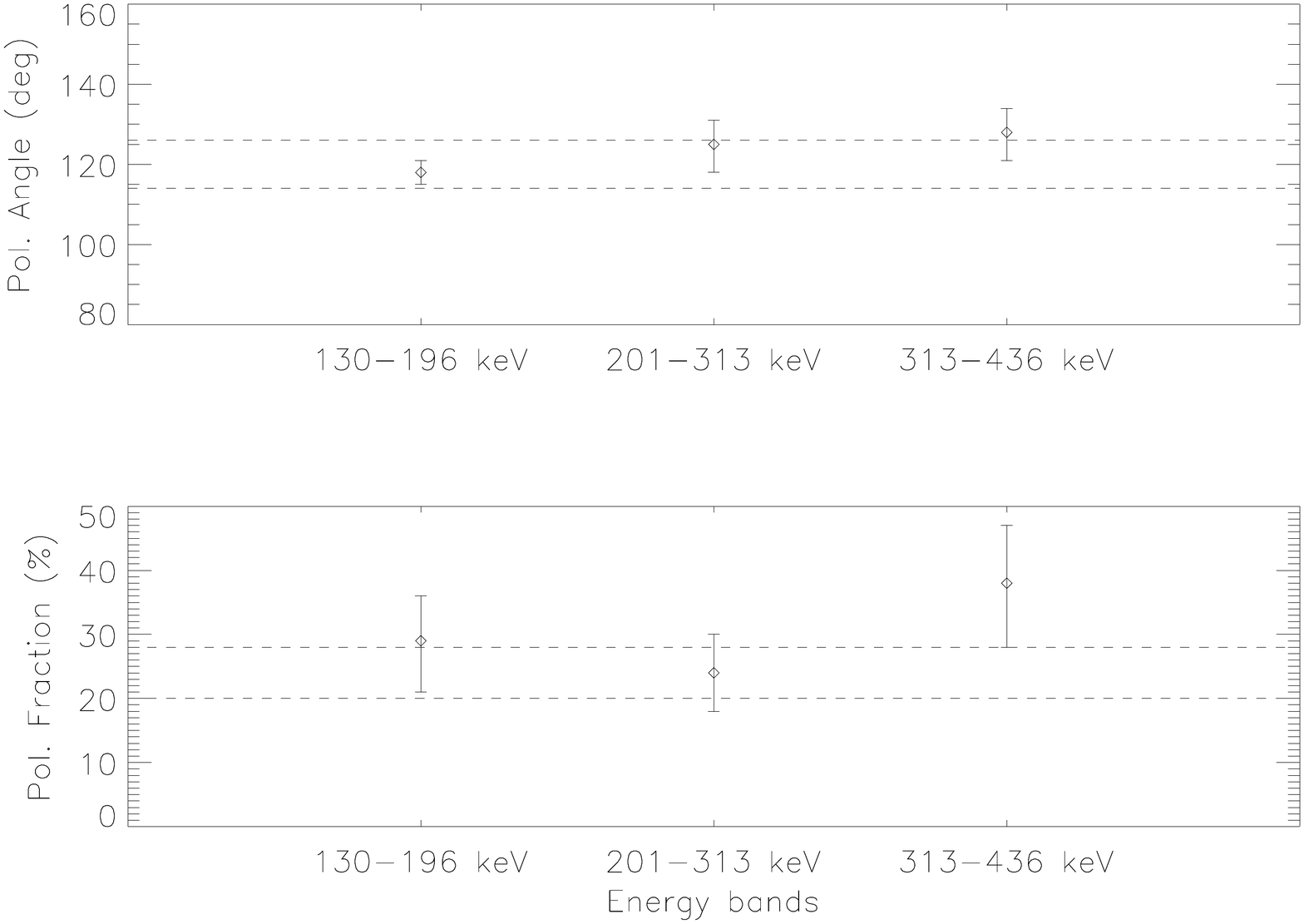}
\caption{Evolution of the polarization parameters with energy for the total data set.
Dashed lines represent the 1-$\sigma$ uncertainty interval for the total data set best-fit values.}\label{polarvsEb}
\end{figure}

To complete our study, we have investigated  the polarization characteristics of the source over energy.
The global energy band (130-436 keV) has been split into 3 energy bins : 130-196 keV, 201-313 keV and 313-436 keV. The polarization parameters have been determined for each bin, for the total data set. The individual  results are displayed in Figure \ref{polarvsEa}, while the evolution of the parameters with energy  is shown in Figure \ref{polarvsEb}, together with  the result from the total energy range as reference (dashed lines = 1 $\sigma$ uncertainty interval). The Polarization Angle  as well as the Polarization Fraction appear, once again, very stable.

\section {Discussion and Conclusion} 
 
 With the SPI spectrometer aboard \textit{INTEGRAL}, we benefit from a large amount of observations in the hard X-ray domain, dedicated to an emblematic source, the Crab nebula. Particularly interesting is the possibility
  to investigate the polarization properties of this source, for the last 16 years.  In the considered  energy range (above 100 keV), the polarimetry studies are less straightforward than in optic or radio.  Concerning SPI, these measurements are based on the Compton interactions of the high energy photons in the detector plane. During a Compton interaction, the polarization  of the incident flux is traced 
 by the angle distribution of the diffused photons. This distribution can be reconstructed on the SPI detector plane, thanks to its 19 individual crystals. Considering the complexity of the polarimetry studies, it is crucial to ensure the reliability of the results, through the reliability of the reconstructed fluxes. This check has been done thanks to the spectral analysis of the  same photons  as those used  in the Polarization analysis. This step testifies that the observed polarized emission comes from the Crab nebula.

Then, we have determined the characteristics of the polarized emission of the Crab nebula, with a robust signal-to-noise ratio, at PA= 120$^\circ \pm 6^\circ$  and PF=24 $\pm 4$\% and established their stability over the 16 years of \textit{INTEGRAL} operations. Moreover, it has been shown that these values do not vary with energy, from 130 to  436 keV. 
Also, they are in good agreement with those obtained with the 2003 observations by \citet{chauvin13},  \textit{INTEGRAL} IBIS \citep{forot08}, PoGOLite+ \citep{chauvin17},  CZTI instrument on AstroSat \citep{vada18} and   SGD on Hitomi \citep{SGD18}, when they consider (as done here)  the total (Pulsar + nebula) Crab  emission. Lastly, our value is comparable to those reported in optical by \citet{optic09}.\\
The polarization properties, even for a source known for its long term stability like the Crab, are most probably variable in space and time: indeed, a more complex behavior appears  clearly as soon as instruments are able to realized spatially or phase resolved analyses (see for instance, \citet{optic09}). However, the global measurements contain the dominant properties  of the source and help to capture a macroscopic picture of this complex region.

The detection of a high  polarization fraction is the definite argument for a synchrotron origin of the hard X-ray emission. Put together with radio and optical studies, this also proves that the same component produces photons from radio to $\sim$ MeV region. 
The unchanged  polarization fraction, deduced from our analysis, reflects the well-known  
  stability of the source. 
Concerning the second parameter, the measured polarization angle corresponds to an electric vector  aligned with the spin axis of the central object  (124$^\circ$, \citet{ng04}) and is also in line with the optical measurements \citep{optic09}. The source steadiness  is still more important in this case, since any variability of the angle weakens or even removes the observable information. Indeed, a variation of the polarization angle smears the angular distribution of the scattered events, thus reducing the observed polarization fraction.

\citet{harding17} have developed a detailed simulation code to reproduce the expected emission from this kind of object, including polarization properties. They consider synchrotron radiation at 
optical to hard X-ray energies and provide phase-averaged and phase-resolved predicted fluxes, polarizations angles and polarization fractions. The predicted values cannot be 
directly compared to observations. However, the expected polarization fractions in the sub-Mev region range  from 10 to 30 \% depending on the assumed geometry, nicely similar to the values deduced from  our observations. This demonstrates 
that  simulation and data analysis works in the polarimetry domain  are in the process of significantly improving our understanding of pulsar physics and high energy photon production in general.

Polarization measurements provide a complementary window particularly valuable to understand the  mechanisms involved in the production of the high energy emission of compact objects. 
The Crab Pulsar and its nebula enjoy a special status in the hard X-ray domain. The stability of the spectral emission, in shape as well as in intensity, is advantageous, particularly for getting high signal-to-noise ratios by accumulating data over long periods, 
or  offering in flight calibration facilities for high energy instruments.  Our results  show that it could serve  as a reference source in the polarimetry domain also, from an instrumental as well as  a modeling point of view.

\section*{Acknowledgments} The \textit{INTEGRAL} SPI project has been completed under the responsibility and leadership of CNES.  We are grateful to ASI, CEA, CNES, DLR, ESA, INTA, NASA and OSTC for support.

\appendix
\section{Tests of Polarization tools}
The SPI polarimetric capacities rely on the Compton interactions of high energy photons in the
detection plane. In the case of a linearly polarized flux, the azimuthal angle distribution of the Compton scattered photons is no longer isotropic. This means that the angular distribution of the diffused photons lays out a specific patterns on the detector plane, directly related to the polarization properties.
Consequently, the polarization analysis relies on the  SPI response, in the specific case of Compton events (or ME for Multiple events).
It is thus important to check that this response is precisely known. This garantees a reliable extraction of the Compton events, and also, permits to identify the spatial distributions  corresponding to different polarized fluxes. \\
The SPI simulations used in the polarization analysis are based on TIMM (for The Integral Mass Model  \citet{TIMM}), translated from GEANT3 into the GEANT4 tool, with further improvements, including the anti-coincidence system configuration and the central mask pixel transparency \citep{chauvin13}.
Each simulation is based on millions of photons, with a parametrable  energy distribution probability (matching the analyzed source spectral shape), and randomly distributed over a large surface to ensure the illumination of the whole instrument. For one photon fired,  all the information are stored (involved  detectors,  energy deposits,...). To finalize a run, the data are processed in the same way as the observational data.\\
We used instrumental data obtained during  the ground calibration campaign, for a mono-energetic (unpolarized) radioactive source at 661 keV, to assess that the Geant4 simulations correctly reproduce the instrument response for both single and multiple detector events. Since the GEANT4 software package \citep{G4} includes the polarization physics (as validated by \citet{G4pol}), we activated this functionality in our code to get simulated count patterns for a set of polarization angles.   
This allows us to check that the spatial distributions of ME are in agreement when considering the unpolarized simulation. Moreover, it demonstrates that a  polarized incident flux results in an anisotropy of the Compton event distribution on the detection plane. For instance, the difference between the unpolarized data mentioned above and a  20$^\circ$ polarized simulation has been evaluated to $\sim 20\%$ (see figures 4 and 5 in \cite{chauvin13}).\\

\begin{deluxetable}{ccccc}
\tablecaption{Observations Log\label{LogP}}
\tablehead{
\colhead{Period} &\colhead{Tstart}&\colhead{Tstop}&\colhead{Useful}&\colhead{included revolutions}\\
\colhead {Number} & &   & \colhead{duration}& 
}
\startdata
P1 & 2003-02-19 & 2003-02-27& 446 ks& 43-44-45 \\
P2 & 2005-10-11 & 2011-10-07 & 1.94 Ms & 365-422-483-541-605-665-665-727-774\\
P2 cont. &   &  &   & 839-902-903-967-968-970-1089-1096\\
P3& 2012-04-10 & 2014-10-06 & 1.82 Ms &1159-1160-1214-1221-1268-1269\\
P3 cont. &  &  &  & 1327-1328-1387-1461-1462 \\
P4 & 2015-03-06 & 2018-09-17 & 2.2 Ms & 1515-1516-1598-1599-1661-1662-1723-1724\\
P4 cont.&   &   &   &1784-1785-1856-1857-1927-1928 -1999-2000\\
\enddata
\end{deluxetable}

\begin{deluxetable}{ccccc}
\tablecaption{Crab nebula best-fit parameters with the Band model. 0.5\% systematic errors included. \label{Fitgrbm}}
\tablehead{
\colhead{Period} &\colhead{$\alpha_1$}&\colhead{$E_{ch}$}&\colhead{$\alpha_2$} &\colhead{$\chi2$ (dof)}\\
%\colhead {} &\colhead{K} & \colhead {kV} & \colhead{} 
}
\startdata
P1  446 ks  & 2.00    & 601 & 2.22   & 77.01   (39) \\
P2  1.94 Ms & 2.01  &  620 & 2.25   & 82.4   (39)\\
P3  1.82 Ms & 2.0 &  602 & 2.32 & 71.9   (39)\\
P4  2.2 Ms & 1.99  & 505  & 2.28 & 85.9 (39)\\
Tot &   2.0    &  572.3   &  2.27   &   351.2 (165)\\    
\enddata
\end{deluxetable}


\begin{thebibliography}{}
%\bibitem[Attie et al. (2003)]{attie03}
%Atti\'e, D., Cordier, B., Gros, M., et al. \ 2003,  \aap, 411, L71
\bibitem[Band et al. (1993)]{band93}
Band, D., Matteson, J., Ford, L., et~al. \ 1993, \apj,  413, 281
%\bibitem[Bühler \& Blandford (2014)]{ReviewCrab}
%B\"uhler, R., \& Blandford, R. \ 2014,  RPPh, 77, 066901
\bibitem[Chauvin et al. (2013)]{chauvin13}
Chauvin, M., Roques, J. P., Clark, D., \& Jourdain, E. \ 2013,  \apj, 769, 137
\bibitem[Chauvin et al. (2017)]{chauvin17}
Chauvin, M.,  Flor\' en, H. -G., Friis, M., et al. \ 2017,   \ Sci. Rep., 7, 7816
\bibitem[Dean et al. (2008)]{dean08}
Dean, A. J., Clark, D. J.,  Stephen, J. B.,  et al. \ 2008, Science, 321, 1183
\bibitem[Forot et al. (2008)]{forot08}
Forot, M., Laurent, P., Grenier, I., et al. \ 2008, \aap, 688, L29
\bibitem[Ferguson et al. (2003)]{TIMM}
Ferguson, C., Barlow, E. J., Bird, A. J. et al. \ 2003, \aap, 411, L19
\bibitem[Geant4 Collaboration (2003)]{G4}
Geant4 Collaboration, Agostinelli, S., Allison, J., et al. \ 2003, \ NIMPA, 506, 250
\bibitem[Harding \& Kalapotharakos (2017)]{harding17}
Harding, A. K., \& Kalapotharakos, C. \ 2017, \apj, 840, 73 
\bibitem[Hitomi collaboration (2019)]{SGD18}
Hitomi Collaboration, Aharonian, F., Akamatsu, H., et al. \ 2018, \ PASJ, 70, 113
\bibitem[Jourdain \& Roques (2009)]{crab09}
Jourdain, E., \& Roques, J. P. \ 2009,  \apj, 704, 17
\bibitem[Mizuno et al. (2005)]{G4pol}
Mizuno, T., Kamae, T., Ng, J. S. T., et al. 2005, NIMPA, 540, 158
\bibitem[Ng \& Romani (2004)]{ng04}
Ng, C.-Y., \& Romani, R. W. \ 2004, \apj,  601, 479
\bibitem[Roques et al. (2003)]{Roques03}
Roques, J. P., Schanne, S., Von Kienlin, A., et~al,\ 2003, \aap, 411, L91
\bibitem[Roques \& Jourdain (2019)]{roques19}
Roques, J. P., \& Jourdain, E. \ 2019,  \apj, 870, 92
\bibitem[S\l owikowska et al. (2009)]{optic09}
S\l owikowska A., Kanbach G., Kramer M., \& Stefanescu A., \ 2009, \mnras, 397, 103
\bibitem[Sturner et al. (2003)]{sturner03}
Sturner S. J.,  Shrader C. R., Weidenspointner G., et~al,\ 2003, \aap, 411, L81 
\bibitem[Vadawale et al. (2018)]{vada18}
Vadawale, S. V., Chattopadhyay, T., Mithun, N. P. S., et al., \ 2018, Nature astron., 2, 50
\bibitem[Vedrenne et al. (2003)]{Vedrenne03}
Vedrenne, G., Roques, J.P., Schonfelder, V., et~al,\ 2003, \aap, 411, L63
%\bibitem[Wilson-Hodge et al.(2011)] {crabvar}
%Wilson-Hodge, C. A., Cherry, M. L.,Case, G. L., et ~al. \ 2011, \apj, 727, L40
\end{thebibliography}
\end{document}